# A Study of Grid Applications: Scheduling Perspective

Aleksandar Lazarević, Dr. Lionel Sacks

Advanced Systems Engineering Group, Dept. of Electronic and Electrical Engineering, University College London

{a.lazarevic, l.sacks}@ee.ucl.ac.uk

**Abstract:** As the Grid evolves from a high performance cluster middleware to a multipurpose utility computing framework, a good understanding of Grid applications, their statistics and utilisation patterns is required. This study looks at job execution times and resource utilisations in a Grid environment, and their significance in cluster and network dimensioning, local level scheduling and resource management.

## 1 Introduction

Utility computing, the concept of provisioning computational resources remotely as a service charged on a per-usage basis, has strong proponents in the industry and scientific community alike. Main driver for the adoption of this approach is its potential to greatly reduce total cost of ownership, provide better value through economies of scale, and offer a virtually unlimited, burstable capacity through dynamic resource syndication.

Grid computing [1], a distributed environment able to integrate computational, storage and network resources across administrative domains, currently provides the scientific community with a high performance computing platform on which a new class of e-Science has been developed. But as it often happens in technology, today's leading edge will become tomorrow's main stream and Grid technology is well positioned to become a leading platform for utility computing.

Such move would significantly change the application set running on the Grid; away from extremely intensive specialised software to a wider ecosystem of medium to long running off-the-shelf applications. In order to properly dimension the computational "power plant" and associated access network of this future Grid, a good understanding of the Grid applications, their behaviour, resource utilisation and statistics is necessary.

The central premise of the author's work is that the Grid workload is driven by humans and as such is inherently non-random. This work will present a study of Grid application's execution times, exploring their historical trends and the relationships with meta-data associated with each job. If present, a degree of correlation, seasonality or patterning would have significant implications in resource dimensioning and management, and in particular to local level scheduling.

## 2 Grid Applications Data

Obtaining realistic and representative Grid usage data is marred by the lack of standardisation and transparency between different Grid middleware, and the understandable reluctance by the Grid operators to share such sensitive information. Trying to obtain data for a multipurpose, utility-like Grid proved to be difficult, as until recently most Grids have been put together to provide high performance computing for a specific task or a single project. These systems tend to be scheduled through hard partitioning and limiting, and together with the lack of job diversity lead to a skewed overall job statistic.

Data analysis presented in this study was collected in a six month period between July and December 2004 on the University College London Central Computing Cluster (CCC). The facility runs Sun Grid Engine (SGE [2]) middleware, and consists of one hundred dual Xeon machines used by around twenty different e-Science research projects. The workload represents a mix of computationally intensive applications, parallel processes requiring low latency communication and numerous shorter master-slave parameter studies.

The primary motivation for this analysis of historical data was to judge its suitability to support predictive scheduling techniques. In such deadline scheduling scenario a forecast of application's execution time and resource utilisation would be made before it is executed. A schedule would then be created to maximise the chance of the job completing before the requested deadline.

## 3. Observations and Results

In this section, a selection of results with particular significance to Grid scheduling will be presented. Support for job execution deadlines and out-of-order optimisation of schedules depends on the *a priori* knowledge of each job's likely execution time. Our study had looked at both effective CPU time used by a job and the wall-clock execution time. In line with most current Grid systems, SGE allocates one Grid process per CPU. The wall-clock execution time is then the total time a CPU is unavailable to the scheduler, and is a more appropriate metric in the scheduling concept than application raw CPU time.

The analysis is done at two levels. At the higher level, the job meta-data is used to segregate clusters of significantly different behaviour, widely varying job statistics and resource utilisations. The meta-data collected varies between middleware but contains at least the running user's credentials, executable name, parameters and submission time. At the lower level, data points belonging to each behavioural group are treated as a time-series and observed for trends, seasonality, discontinuities and overall distribution.

Although this study was done off-line on historical data, all clustering and analysis methods used have been selected to be suitable for fully automated on-line approach. Ultimate aim would be integration of such passive monitoring and analysis system into the Grid scheduling architecture, and provisioning of a predictive, probabilistic scheduling.

Figure 1. shows a plot of wall-clock execution times for around 45,000 jobs run on the Central Computing Cluster.

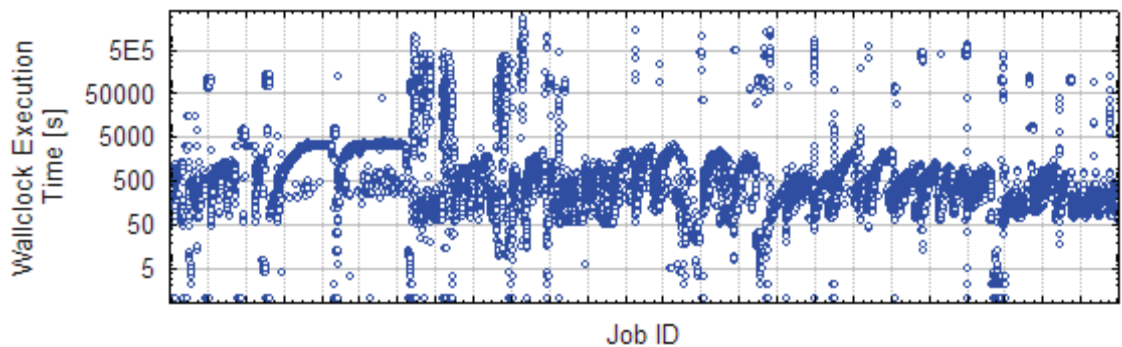

**Figure 1.** Wall-clock execution times for all jobs on CCC Jul '04 – Dec '04

The most striking feature of the plot is a large, six orders of magnitude range of execution times. It is also clear that very short and very long jobs are relatively rare compared to the ones in 50 to 5000 seconds range. Other than a guide to general distribution of run times, little can be inferred from this all-inclusive approach.

By using the job's meta information, the raw data was separated according to the submitting user's Unix group. Assigned by the administrator at the time of account opening, this is effectively the scientific project on which certain users are working. A selection of these groups, and their job execution times are plotted separately, as shown in Figures 2 to 4. These show a significant difference in the distribution of execution times, total number of jobs and the specific workflow.

The execution times exhibit discontinuities and abrupt changes over the given six month period. We treat these events as normal changes in the mode of operation due to the alteration of underlying scientific workflow. At these points in time it is likely that a new data set was introduced, a new scientific objective was identified, application changed or updated, or perhaps the group members have taken on a completely different project. Through robust analysis approach, and corroboration with meta-data, these mode changes can be handled as normal stream of operation rather then an exception. An example of a mode change is show in Figure 5. Execution times initially follow a linear upward trend reaching around 650 seconds before dropping to sub-200 seconds range with no trend.

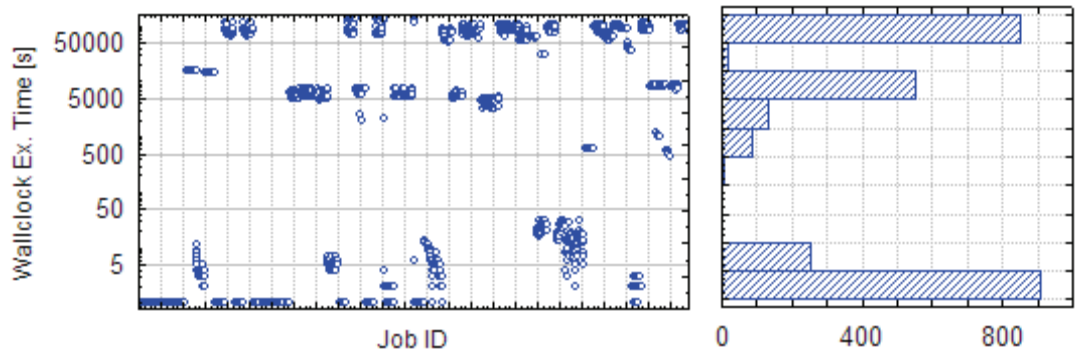

**Figure 2.** Execution times plot and histogram for *geogrs* job group

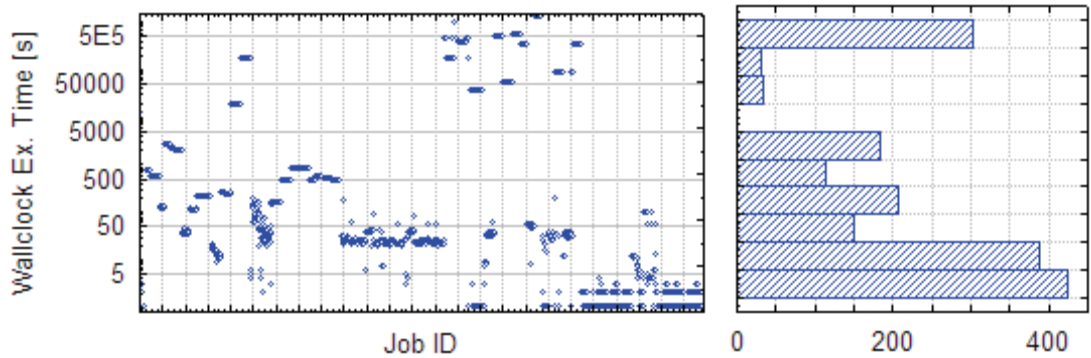

**Figure 3.** Execution times plot and histogram for *chemccs* job group

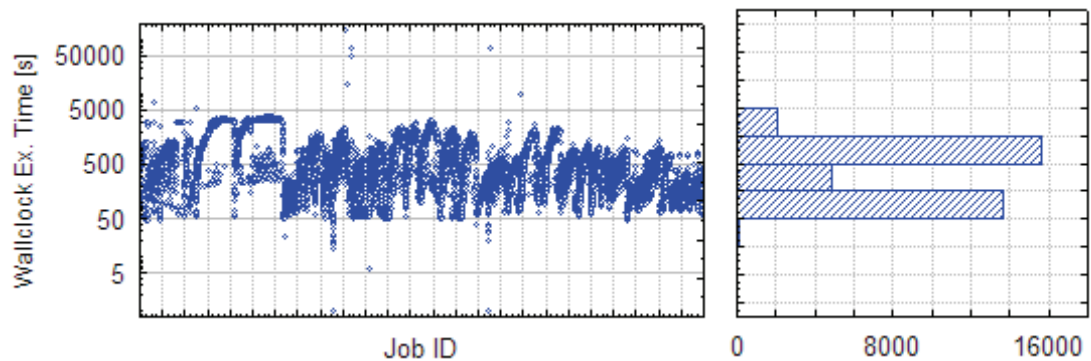

**Figure 4.** Execution times plot and histogram for *matsim* job group

In between mode changes, execution times are usually clustered, with significant short to medium term autocorrelation. They exhibit random fluctuations either around a mean value, or around a linear or exponential trend. Such behaviour is significant from the perspective of predictive scheduling: providing a robust approach is taken at recognising mode changes and discontinuities, a degree of predictability exists in the "well-behaved" region.

At the highest level of granularity, short, sudden, isolated and significant changes in the observed values are classified as anomalies. These can occur for a number of reasons, such as short network connectivity problems or spikes in utilisation, inconsistencies in the application's data set or simply due to system crashes, lockups or monitoring problems. Figure 5. shows several of these outliers on both sides of the discontinuity. Despite their treatment as an irregularity, anomalies are an always present feature of a highly probabilistic data such as job execution times and can carry a significant information value. For example an increased frequency of anomalies may indicate an onset of a mode change for the whole group. Their value in making good forecasts will be investigated further.

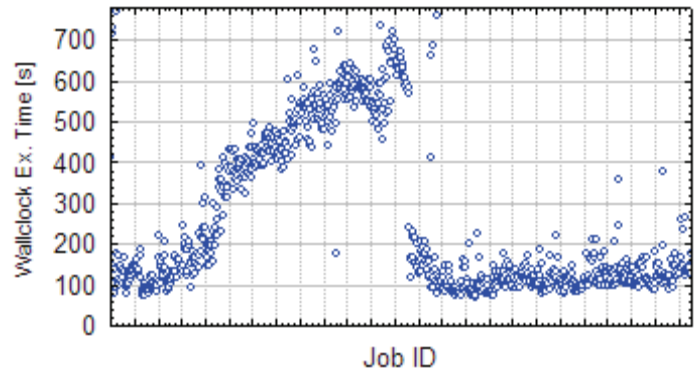

Figure 5. Execution Time Discontinuity - matsim Group

## 4. Conclusions and Future Work

The paper introduces a study of Grid applications, their resource utilisation, statistical properties and correlation with historical values on a typical multi-purpose Grid cluster. The work aims to motivate better use of monitoring and accounting data already routinely collected on these systems. By means of a suitable analysis and correlation between meta- and historical data, valuable information can be fed back into the system and used as an input to an automated predictive scheduling system.

The study has established that overall Grid workload is made up of a number of statistically different groups of users, applications and modes of operation. Segregation of these could be done using clustering techniques on the meta-data maintained about the job, and predictions are possible given a high degree of autocorrelation within a job's single continuous mode of operation. However, due to a highly dynamic nature of the Grid workload, a flexible and agnostic approach throughout the whole system is needed. Changes in statistical distributions, absolute values and clustering properties are ongoing and likely to happen at different time scales for different job groups. Values for anomaly detection thresholds, mode changes and trend analysis should be dynamically set and within the context of the job's current state. A truly probabilistic approach from the ground up would also enable better handling of unavoidable forecast errors.

Primary concern in analysing Grid usage is the level to which the data represents a genuine Grid scenario. CCC environment is a production system running on popular middleware, serving a number of different real-life scientific projects. As such, it is representative of the likely current target segment for a utility cluster based on Grid technology. Although the specific resource utilisation distributions, or meta-data clustering methods, may not be globally applicable we believe that the idea of analysing the usage trends and actively using them in job scheduling remains valid.

The study opens many avenues for further exploration. Integration of historical usage information, current monitoring data and predictions of future expected load in a timely way and with a known degree of accuracy can be invaluable in managing the system, dimensioning the access network and the computational pool as well as supporting advanced Quality of Service and Service Level Agreements. Our research will focus on utilising models developed on the historical data to support predictive and probabilistic Grid scheduling.